\documentclass[12pt]{iopart}
\usepackage{iopams}

\newtheorem{lemma}{Lemma}

\newcommand{\proof } {\noindent{\bf{Proof.}} }

\begin{document}
\today

\title{Analysis of SM quantum information}

\author{C J O'Loan}

\address{School of Mathematics and Statistics, University of St Andrews, 
 KY16~9SS, UK}

\ead{cjo2@st-and.ac.uk}

\begin{abstract}
Morozova and Chentsov \cite{moro90} studied Riemannian metrics on the set of probability measures. They showed that, up to a constant factor, the Fisher information is the only Riemannian metric which is monotone under stochastic transformation.  Sarovar and Milburn \cite{saromilb06} computed an upper bound on the Fisher information for one-parameter channels. In \cite{oloan07} we extended their bound to an upper bound on the Fisher information of multi-parameter families of states; we call this the SM quantum information.  Petz and Sud\'ar  \cite{petz95} characterized fully the set of monotone metrics on the space of all density matrices. We analyse the SM quantum information in light of their work. We show that the SM quantum information is not a well-defined metric on the space of density matrices: different choices of phase of the eigenvectors lead to different metrics. We define a new metric $C_L$ as a lower bound among the SM quantum informations. We look at properties of $C_L$ and show that it is invariant but not monotone.

\end{abstract}
\submitto{\JPA}
\maketitle
 
\section{Introduction} \label{sec:intro}
\subsection{Metrics}
Metrics are important. 
In statistics the Fisher information is the unique monotone metric on the space of probability measures \cite{moro90}. It tells us how precise an estimate is and gives us confidence intervals. Often we may be interested in estimating a parameter from a given distribution. The Fisher information tells us how precise our estimate is, i.e.\ how confident we can be that our estimate is close to the true value.

There is no unique monotone metric on the space  of all density matrices  \cite{petz95}. There are a few used in recent literature which we will introduce later. When we are estimating a quantum state, the metric gives us a measure of how precise our estimate is. As state estimation is fundamental to the field of quantum information, the study of metrics on the space of quantum states is of importance.

 We will now look briefly at the theory of one-parameter quantum estimation. Quantum estimation is concerned with estimating (especially optimally) quantum states and processes. Two important mathematical objects in quantum estimation are density matrices and  positive operator-valued measures (POVMs). A density matrix represents the state of the quantum system. The density matrix of a $d$-dimensional state is a $d \times d$ non-negative, Hermitian matrix, with trace $1$, i.e.
 \begin{eqnarray*}
 \langle v | \rho | v \rangle \geq 0,\quad  \forall | v \rangle, \quad \rho^{\dagger} = \rho , \quad \tr\{\rho\} =1.
 \end{eqnarray*}
 A POVM is represented by a set of operators $\{ M_m \}$ which are Hermitian, non-negative and sum to the identity, i.e.
  \begin{eqnarray*}
M_m^{\dagger} =  M_m , \quad \langle v | M_m | v \rangle \geq 0,\quad  \forall | v \rangle,  \quad \sum_m M_m = \mathbb{I}.
 \end{eqnarray*}
 Given a state $\rho(\theta)$ with an unknown parameter $\theta$ and a POVM $\{ M_m \}$, the probability density of the measurement yielding the result $x_m$ is
 \begin{eqnarray*}
p(x_m ; \theta) = \tr \{  \rho(\theta)  M_m \} .
\label{eq:born}
\end{eqnarray*}
These probabilities usually depend on $\theta$.   In practice, we repeat a measurement $N$ times on identical copies of the state to be estimated.  The outcomes of the measurement depend probabilistically on the parameter $\theta$. We then choose an estimator which gives us an estimate of $\theta$ from the measurement results. 
For optimal estimation of a state, we choose the POVM and estimator that give us the most `information' about the state. 

A standard way of quantifying the performances of input states and POVMs is to use Fisher information. Intuitively, Fisher information tells us the amount of `information' about $\theta$ contained in a measurement result. Fisher information is defined as
\begin{eqnarray*}
 F_{M}(\theta) &\equiv&  \int   p(\xi;\theta) \left( \frac{\partial \ln p(\xi;\theta)}{\partial \theta} \right)^2 d\xi \\
&=&  \int  \frac{1}{p(\xi;\theta)} \left( \frac{\partial p(\xi;\theta)}{\partial \theta} \right)^2 d\xi.\\
\label{eq:FI00}
\end{eqnarray*}
If the measurement outcomes are discrete with probabilities $p_1(\theta),\dots ,p_n(\theta)$, then the Fisher information can be expressed as
\begin{eqnarray*}
 F_{M}(\theta) = \sum_{k=1}^{n} \frac{1}{p_k (\theta)} \left( \frac{d p_k (\theta)}{d \theta}\right)^2 .
\label{eq:FI01}
\end{eqnarray*}
The importance of Fisher information is seen in the Cram\'er--Rao inequality. This states that the variance of an unbiased estimator $t$ is greater than or equal to the reciprocal of the Fisher information, i.e.
\begin{equation*}
\mathrm{var}_{\theta}[t(x)] \geq \frac{1}{ F_{M}(\theta)}.
\label{eq:CRineq}
\end{equation*}
Under mild regularity conditions for $p(x ; \theta)$, a maximum likelihood estimator achieves this lower bound asymptotically, see section 8.9 of \cite{vaart98}. The larger the Fisher information, the more accurately we can estimate the unknown parameter. A standard approach to estimation is to choose the procedure  which maximises the Fisher information and use the maximum-likelihood estimator.

When we look at metrics on the space of quantum states the situation is more complex than on the space of probability measures. Petz and Sud\'ar \cite{petz95} showed that there is no unique monotone metric  on the space of quantum states. They characterized fully the set of monotone metrics. 
The most frequently encountered monotone metrics in recent literature are the {\it Symmetric Logarithmic Derivative (SLD)}, {\it Kubo-Mori Bogoliubov (KMB)} and {\it Right Logarithmic Derivative (RLD)} metrics.
The SLD is the most widely used, as it is the smallest among the set of monotone metrics on the space of all quantum states \cite{petz95}.

The SLD quantum information $H_{SLD}(\theta)$  upper bounds the Fisher information, i.e.
\begin{equation}
F_M(\theta) \leq H_{SLD}(\theta).
\label{eq:fleqh}
\end{equation}
For one-parameter models it is achievable. The SLD quantum information has been used widely in the estimation of states \cite{helstrom76,helstrom76b,holevo82,hayashi05} and quantum channels \cite{fuj01a,fuj01,fujimai03,fuj04,ballestera,ballesterb}. 
The SLD quantum information is defined as
\begin{equation*}
H_{SLD}(\theta) = \tr \{ \rho(\theta) \lambda(\theta)^2 \},
\end{equation*}
where $\lambda(\theta)$ is the SLD quantum score, defined as any self-adjoint solution to the matrix equation
\begin{equation*}
\frac{d \rho(\theta)}{d \theta} = \frac{1}{2}(\rho(\theta) \lambda(\theta) + \lambda(\theta) \rho(\theta)).
\label{eq.sld}
\end{equation*}
Equality holds in (\ref{eq:fleqh}) if and only if the POVM $\{ M_m \}$ satisfies
\begin{equation}
M_x^{(1/2)} \lambda(\theta) \rho^{(1/2)}(\theta) = \xi_x(\theta)  M_x^{(1/2)} \rho^{(1/2)}(\theta), \quad \forall  x,
\label{eq:equalcondfh}
\end{equation}
where $\xi_x(\theta)$ is a real number.
Since the SLD is hermitian, it can be diagonalized, i.e. written in the form
\begin{equation*}
\lambda(\theta) = \sum_i \mu_i(\theta) | e_i(\theta) \rangle \langle e_i(\theta)|.
\label{eq.slddiag}
\end{equation*}  
For the one-parameter family of states $\rho(\theta)$ the POVM $\{ | e_i(\theta) \rangle \langle e_i(\theta)|\}$ satisfies (\ref{eq:equalcondfh}) and hence is optimal at $\theta$. The resulting Fisher information equals the SLD quantum information.

The KMB quantum information $H(\theta)_{KMB}$ for the family of states $\rho(\theta)$ is defined as
\begin{equation*}
H_{KMB}(\theta) = \tr\left\{\rho(\theta) \left(\frac{d \log \rho(\theta)}{d \theta}\right)^2 \right\},
\end{equation*}
providing that $\rho(\theta)$ has full rank. The `classical' Fisher information is the limit of the `classical' relative entropy $D(p \| q) = \sum_{i=1}^k p_i \ln (p_i/q_i)$. That is, given a probability simplex $p_\theta = \{ p_i(\theta) \}$,  
\begin{equation*}
F_M(\theta) = \lim_{\epsilon \rightarrow 0} \frac{2 D(p_{\theta} \| p_{\theta + \epsilon} )}{\epsilon^2}.
\end{equation*}
  Similarly, the KMB quantum information has been shown to be equal to the limit of the quantum relative entropy $D(\rho \| \sigma ) = \tr ( \rho ( \ln \rho - \ln \sigma))$ \cite{hayashi02}. That is,
\begin{equation*}
H_{KMB}(\theta) = \lim_{\epsilon \rightarrow 0} \frac{2 D(\rho_{\theta} \| \rho_{\theta + \epsilon} )}{\epsilon^2}.
\end{equation*}
From an information-geometrical point of view the KMB is the most natural quantum extension of Fisher information  \cite{hayashi02}.

The RLD quantum information for the family of states $\rho(\theta)$ is defined as
\begin{equation*}
H_{RLD}(\theta) = \tr\left\{\rho(\theta)^{-1} \left(\frac{d \rho(\theta)}{d \theta}\right)^2 \right\},
\end{equation*}
providing that $\rho(\theta)$ has full rank. It is the maximal metric among monotone metrics on the space of all density matrices  \cite{petz95}. The RLD has also been used in estimation theory \cite{fuji94}. 

\subsection{The SM quantum information}
The SM quantum information was introduced by Sarovar and Milburn \cite{saromilb06}. It is a conveniently computable upper bound on the Fisher information for one-parameter quantum channels of the form
\begin{equation*}
\rho_0 \mapsto \sum_k E_k(\theta) \rho_0 E_k(\theta)^\dagger.
\end{equation*}
Sarovar and Milburn's bound is defined as
\begin{equation*}
C_{\Upsilon}(\theta) = 4 \sum_k \tr \{ \Upsilon_k(\theta)' \rho_0 \Upsilon_k(\theta)^{' \dagger} \}, \qquad  \Upsilon_k'(\theta) = \frac{d}{d\theta}  \Upsilon_k(\theta),
\end{equation*}
where $\Upsilon_k(\theta)$ are the canonical Kraus operators, which are uniquely defined, up to a phase, as the operators satisfying $\tr\{\Upsilon_j(\theta) \rho_0 \Upsilon_k(\theta)^\dagger \} = \delta_{jk} p_k(\theta)$ for all $j,k$.
These operators are used to avoid ambiguity in $C_\Upsilon$, since there is no unique Kraus decomposition for a quantum channel, see p. 370 of \cite{chuang00}.

In \cite{oloan07} we showed that the Riemannian metric $C_{\Upsilon}(\theta)$ on a one-parameter family of channels $\mathcal{E}_\theta$, could be extended to a Riemannian metric on a parametric family of states $\rho_\theta$.

\subsection{Definition of Terms}
We investigate whether $C_{\Upsilon}(\theta)$ is a well-defined metric. 
By `well-defined metric' we mean that the metric is invariant and monotone.

{\noindent{ \em Invariance.}} 
Given two parametric families of states $\rho_\theta$ and $\sigma_{\theta}$ we say that $\rho_\theta$ and $\sigma_{\theta}$ are {\it equivalent}, $\rho_\theta \sim \sigma_{\theta}$, if there exist two fixed {\it Trace Preserving, Completely Positive (TP-CP) maps} $\mathcal{E}, \mathcal{F}$ such that 
\begin{equation*}
\rho_\theta= \mathcal{E}(\sigma_{\theta}), \qquad \sigma_{\theta} = \mathcal{F}(\rho_\theta).
\end{equation*}
We say that the metric $J$ is {\it invariant} if
\begin{equation*}
\rho_\theta \sim \sigma_{\theta} \quad \mathrm{implies} \quad  J(\rho_\theta) = J(\sigma_{\theta}).
\end{equation*}
{\noindent{\em Monotonicity.}} 
We say that the metric $J$ is {\it monotone} if 
\begin{equation*}
J(\rho_\theta) \geq J(\mathcal{E}(\rho_\theta))
\end{equation*}
for all TP-CP maps $\mathcal{E}$.

Invariance is an essential property for a meaningful metric; monotonicity is a desirable but not essential property.
We show that $C_{\Upsilon}(\theta)$ is not a well-defined metric. 
Starting from $C_\Upsilon(\theta)$, we define a new Riemannian metric $C_L(\theta)$. We show that $C_L(\theta)$ is invariant but not monotone.
We show how $C_L(\theta)$ is related to $C_\Upsilon(\theta)$ and to the SLD quantum information $H(\theta)$.

Morozova and Chentsov \cite{moro90} showed that for any invariant Riemannian metric on the space of all density matrices,  at a particular density matrix, represented in a certain basis as $\rho = \mathrm{diag}(p_1, \dots, p_n)$, the squared length of any tangent vector $A = (A_{jk})$  must be of the form
\begin{equation}
C \sum_i \frac{1}{p_i} | A_{ii} |^2 + 2 \sum_{j<k} c(p_j, p_k)  | A_{jk}|^2,
\label{eq:chemor1}
\end{equation}
where $C$ is a constant, the function $c(x, y)$ is symmetric, i.e.\ $c(x,y) = c(y,x)$, and $c(\lambda x,  \lambda y) = \lambda^{-1} c(x, y)$. We are interested in the distance $d ( \rho, \rho + A)$.
When we are looking at parametric families of states, we are interested in the distance between nearby states $\rho_\theta$ and $\rho_{\theta+\epsilon}$, i.e. we take
\begin{equation*}
d ( \rho_\theta, \rho_{\theta+\epsilon} ) \approx d ( \rho_\theta, \rho_{\theta} + \epsilon \frac{d \rho_\theta}{d \theta}) .
\end{equation*}
In this case any invariant Riemannian metric must be of the form
\begin{equation}
C \sum_i \frac{1}{p_i} \left| \frac{d \rho}{d \theta}_{ii} \right|^2 + 2 \sum_{j<k} c(p_j, p_k)  \left| \frac{d \rho}{d \theta}_{jk}\right|^2.
\label{eq:chemor}
\end{equation}

Petz and Sud\'ar \cite{petz95} showed that any monotone Riemannian metric on the space of all density matrices must be of the form (\ref{eq:chemor1}) and that the function $f(t) = 1/c(t,1)$ must be monotone.
 For the SLD , KMB  and RLD quantum informations  \cite{petz95} , $C =1$ and
\begin{eqnarray*}
c_{SLD}(x,y) &=& \frac{2}{x + y} \\
c_{KMB}(x,y)  &=& \frac{\ln x - \ln y}{x - y} \\
c_{RLD}(x,y)  &=& \frac{1}{2}\bigg(\frac{1}{x} + \frac{1}{y}\bigg).
\end{eqnarray*}

\section{Analysis of the SM quantum information}
 The SM quantum information was originally defined \cite{saromilb06} as an upper bound on the Fisher information obtained from one-parameter families of channels. In \cite{oloan07} we showed that the SM quantum information puts an upper bound on the Fisher information for a family of states, and hence is a Riemannian metric on the space of a parametric family of states. The SM quantum information for the family of states
 \begin{equation}
 \rho(\theta) = \sum_{k=1}^d p_k(\theta) | w_k (\theta) \rangle \langle w_k(\theta) |,
\label{eq.statesw}
 \end{equation}
was shown in \cite{oloan07} to be equal to
  \begin{eqnarray}
C_{\Upsilon}(\theta) =  \sum_i \frac{1}{p_i(\theta)} \bigg(\frac{d p_i}{d \theta} \bigg)^2 &+& 4 \sum_{j<k} (p_j(\theta) + p_k(\theta)) | \langle w_j' | w_k\rangle |^2 \nonumber  \\
  &+& 4 \sum_i p_i(\theta) | \langle w_i' | w_i \rangle |^2.
   \label{eq.origsm}
 \end{eqnarray}
  We can rewrite this as 
    \begin{eqnarray*}
C_{\Upsilon}(\theta) =    \sum_i \frac{1}{p_i(\theta)} \bigg(\frac{d p_i}{d \theta} \bigg)^2 &+& 4 \sum_{j<k} \frac{p_j(\theta) + p_k(\theta)}{(p_j(\theta) - p_k(\theta))^2}  \bigg| \bigg\langle w_j \bigg| \frac{d \rho}{d \theta} \bigg| w_k \bigg\rangle \bigg|^2 \nonumber  \\
  &+& 4 \sum_i p_i(\theta) | \langle w_i' | w_i \rangle |^2. 
\label{eq:smche}
\end{eqnarray*}
We see that $C_{\Upsilon}(\theta)$ not of the form (\ref{eq:chemor}), and hence is neither invariant nor monotone. The SM quantum information $C_{\Upsilon}(\theta)$ for a family of states is defined in terms of its eigenvectors and eigenvalues (\ref{eq.origsm}) \cite{oloan07}. The eigenvectors of a state are unique up to a change of phase. It turns out that different choices of phase for the eigenvectors lead to different metrics.

{\noindent{\bf{Example 1}} 
\newline
Consider the set of 2 dimensional states,
\begin{equation*}
\rho(r,\theta,\phi) =  \frac{1}{2} \left( \begin{array}{cc}
\ 1 + r \cos(\theta) & r \sin(\theta) \exp( - i \phi) \\
\ r \sin(\theta) \exp( i \phi)  & 1 -r \cos(\theta)   \end{array} \right).
\end{equation*}
This has spectral decomposition
\begin{eqnarray*}
\rho(r,\theta,\phi) &=&\frac{1+r}{2} \bigg| v_1 (\theta,\phi) \bigg \rangle \bigg \langle v_1 (\theta,\phi) \bigg | + \frac{1-r}{2} \bigg | v_2 (\theta,\phi) \bigg \rangle \bigg \langle v_2 (\theta,\phi) \bigg | ,\\
| v_1 (\theta,\phi)\rangle &=& ( \cos(\theta/2) \exp(-i \phi/2), \sin(\theta/2 ) \exp(i \phi/2))^T,\\
| v_2 (\theta,\phi)\rangle &=& ( \sin(\theta/2) \exp(-i \phi/2), - \cos(\theta/2) \exp(i \phi/2))^T.
\end{eqnarray*}
 The SM quantum information for the family of states $\rho(\theta)$ calculated from the above eigenvalues and eigenvectors is 
  \begin{equation*}
 H_{SM} = \left( \begin{array}{ccc}
\ \displaystyle \frac{1}{1-r^2} & 0 & 0\\
\ 0 & 1 & 0 \\
\ 0 & 0 &1  \end{array} \right).
\end{equation*}
 If we change the eigenvectors by a phase shift, i.e. $| v_k(\theta,\phi) \rangle \mapsto \exp(-i \phi/2)| v_k(\theta,\phi) \rangle$, then the density matrix is unchanged, but the SM quantum information calculated from the eigenvalues and shifted eigenvectors  becomes
   \begin{equation*}
 H_{SM} = \left( \begin{array}{ccc}
\ \displaystyle \frac{1}{1-r^2} & 0 & 0\\
\ 0 & 1 & 0 \\
\ 0 & 0 &2 + 2r \cos\theta  \end{array} \right).
\end{equation*}
Hence the SM quantum information is not a meaningful metric on the space of all density matrices.

\section{A new metric}
 We define the  $C_L$ quantum information for the family of states (\ref{eq.statesw}) as
\begin{eqnarray}
C_L &=& C_{\Upsilon} - 4 \sum_i p_i | \langle w_i' | w_i \rangle |^2 \label{eq.dercl} \\
&=& \sum_i \frac{1}{p_i} \bigg(\frac{d p_i}{d \theta} \bigg)^2 + 4 \sum_{j<k} (p_j + p_k)  | \langle w_j' | w_k \rangle|^2 \nonumber \\
&=& \sum_i \frac{1}{p_i} \bigg(\frac{d p_i}{d \theta} \bigg)^2 + 4 \sum_{j<k} \frac{p_j + p_k}{(p_j - p_k)^2}  \bigg| \bigg\langle w_j \bigg| \frac{d \rho}{d \theta} \bigg| w_k \bigg\rangle \bigg|^2  \nonumber. 
  \label{calebmetric}
\end{eqnarray}
The $C_L$ quantum information is of the form (\ref{eq:chemor}) with $C=1$ and
\begin{equation*}
c_{L} = 2 \frac{p_j + p_k}{(p_j - p_k)^2}.
\end{equation*}
This function is symmetric and  $c_L(\lambda x,  \lambda y) = \lambda^{-1} c_L(x, y)$. Hence, $C_L$ is an invariant metric on the set of parameterised channels $\rho(\theta)$. As it is invariant, $C_L$ does not suffer the same defect as $C_\Upsilon$: a family of states $\rho(\theta)$ leads to a unique metric.
Using the function $c_L$ and putting $C=1$, we can write $C_L$ in the form (\ref{eq:chemor1}) and hence extend $C_L$ to be a meaningful metric on the space of all density matrices. 

For a metric to be monotone, it must be of the form (\ref{eq:chemor1}) and the function $f(t)$ associated with the metric must be monotone and satisfy $f(t) = tf(t^{-1})$.
  The functions associated with the SLD, KMB and RLD quantum informations are
  \begin{eqnarray*}
  f_{SLD}(t) &=& \frac{1+t}{2}\\
  f_{KMB}(t) &=& \frac{t-1}{\log t}\\
  f_{RLD}(t) &=& \frac{2t}{1+t}.
  \end{eqnarray*}
 The function associated with $C_L$ is
 \begin{equation*}
 f_{C_L}(t) = \frac{(t -1)^2}{2(1+t)}.
 \end{equation*}
If $f$ is a monotone function then $f(0) \leq f(t_1) \leq f(t_2)$ whenever $0 \leq t_1 \leq t_2$. The function $f_{C_L}(t)$ satisfies $f_{C_L}(t) = tf_{c_L}(t^{-1})$ but is not monotone, as $ f_{C_L}(0) > f_{C_L}(1)$. Hence, $C_L$ is an invariant but not monotone Riemannian metric.
 
{\noindent{\bf{Example 2}} 
\newline
The {\it depolarizing channel}, see p. 378 of \cite{chuang00}, acts on states in the following way
\begin{equation*}
\rho_0 \mapsto r \rho_0 + \frac{1-r}{d} \mathbb{I}, \qquad \frac{-1}{d^2-1} \leq r \leq 1.
\end{equation*}
Consider the one-parameter set of mixed states 
  \begin{eqnarray*}
\rho(\theta) &=& (1-2 \epsilon) | v_1 \rangle \langle v_1 | + \epsilon | v_2 \rangle \langle v_2 | + \epsilon | v_3 \rangle \langle v_3 | ,\\
| v_1 \rangle &=& ( 1, 0, 0 )^T,\\
| v_2 \rangle &=& ( 0, \cos\theta, \sin\theta)^T,\\
| v_3 \rangle &=& ( 0, -\sin\theta, \cos\theta)^T,
\end{eqnarray*}
where $\epsilon$ is fixed. The $C_L(\theta)$ quantum information of this family of states is $8 \epsilon$.
Under the action of the depolarizing channel we get 
\begin{eqnarray*}
\mathcal{E}(\rho(\theta)) &=& \bigg( r(1-2 \epsilon) + \frac{1-r}{3}\bigg) | v_1 \rangle \langle v_1 | \\
&+& \bigg( r \epsilon + \frac{1-r}{3}\bigg) | v_2 \rangle \langle v_2 | + \bigg( r \epsilon + \frac{1-r}{3}\bigg) | v_3 \rangle \langle v_3 | 
\end{eqnarray*}
with $| v_i \rangle$ unchanged. The $C_L$ quantum information for the family of states $\mathcal{E}(\rho(\theta))$ is $8 r \epsilon + 8(1-r)/3$. Now
\begin{equation*}
C_{L}(\mathcal{E}(\rho(\theta))) - C_{L}(\rho(\theta)) = (1-r) \bigg( \frac{8}{3} - 8 \epsilon \bigg).
\end{equation*}
For $r < 1$ and $\epsilon < 1/3$,  $C_L$ has increased under the action of a TP-CP map, thus demonstrating the non-monotonicity of $C_L$.

\section{Relationships between $C_L, C_\Upsilon$ and $H$}
In this section we prove the following relationships,
\begin{equation*}
H_{SLD}(\theta) \leq C_L(\theta) \leq C_\Upsilon(\theta).
\end{equation*} 
We show that the above inequalities hold for the one-parameter and multi-parameter cases.

\subsection{One-parameter case }
\begin{lemma}
\begin{eqnarray}
C_L(\theta) \leq C_{\Upsilon}(\theta)
\label{eq.clecups}
\end{eqnarray}
with equality if and only if
\begin{eqnarray}
 \sum_i p_i | \langle w_i' | w_i \rangle |^2 = 0.
 \label{eq.eq1pclcu}
 \end{eqnarray}
\label{lem.clleqcups1}
\end{lemma}
\proof
This follows from the definition of $C_L$ (\ref{eq.dercl}) and the fact that $ 4 \sum_i p_i | \langle w_i' | w_i \rangle |^2$  is non-negative.

\begin{lemma}
The quantity $C_L$ is an upper bound on the SLD quantum information, i.e.
\begin{equation*}
H_{SLD}(\theta) \leq C_L(\theta).
\label{eq.1phleqcl}
\end{equation*}
\label{lem.clleqh}
\end{lemma}
\proof
From \cite{oloan07}
\begin{equation*}
H_{SLD}(\theta) = \sum_{k} \frac{1}{p_k} \bigg(\frac{d p_k}{d \theta} \bigg)^2 +   \sum_{  j < k} 4\frac{ ( p_{j} - p_{k})^2}{p_{j} + p_{k}} |\langle w_{j}' | w_{k}  \rangle|^2,
\label{eq:Hsldr}
\end{equation*}
and hence
\begin{equation}
C_L(\theta) - H_{SLD}(\theta) = 16 \sum_{  j < k } \frac{p_{j} p_{k}}{p_{j} + p_{k}} |\langle w_{j}' | w_{k}  \rangle|^2 = 8 \sum_{  j \neq k } \frac{p_{j} p_{k}}{p_{j} + p_{k}} |\langle w_{j}' | w_{k}  \rangle|^2, \quad 
\label{eq.clminush}
\end{equation}
since $|\langle w_{j}' | w_{k}  \rangle|^2$ is symmetric with respect to $j$ and $k$ \cite{oloan07}.
The right hand side of (\ref{eq.clminush}) is non-negative and hence the result follows.
\begin{lemma}
Equality holds in 
\begin{equation*}
H_{SLD}(\theta) \leq C_L(\theta)
\end{equation*}
if and only if 
 \begin{equation*}
\langle w_{j}' | w_{k}  \rangle = 0, \quad \forall j \neq k, p_j, p_k  >0.
\end{equation*}
\label{eq.eq1phcl}
 \end{lemma}
 \proof
 This follows from the right hand side of (\ref{eq.clminush}).
 \newline
  \newline
{\noindent{\bf{Remark}} 
\newline
Unlike some other invariant quantum informations such as the RLD and KMB, which put upper bounds on the SLD quantum information, the $C_L$ quantum information is defined for pure states. For pure states $p_j p_k = 0$ for all $j \neq k$, so the right hand side of (\ref{eq.clminush}) is zero and we have $C_L(\rho_\theta) = H_{SLD}(\rho_\theta)$.

\subsection{The multi-parameter case}
In the multi-parameter case the SM quantum information is the matrix with entries
\begin{eqnarray*}
C_{\Upsilon}(\theta)_{kl} &=& \sum_i \frac{1}{p_i} \bigg(\frac{\partial p_i}{\partial \theta^k} \bigg)\bigg(\frac{\partial p_i}{\partial \theta^l} \bigg)  + 4 \Re \sum_{i < j} (p_i + p_j ) \bigg\langle w_i^{(k)} \bigg| w_j \bigg\rangle \bigg\langle w_j \bigg| w_i^{(l)} \bigg\rangle \nonumber \\
&+& 4  \sum_i p_i \bigg\langle w_i^{(k)} \bigg| w_i \bigg\rangle \bigg\langle w_i \bigg| w_i^{(l)} \bigg\rangle.
\label{eq.multicups}
\end{eqnarray*}
We define the multivariate version of $C_L$ as the matrix with entries
\begin{eqnarray*}
C_L(\theta)_{kl} &=& C_{\Upsilon}(\theta)_{kl}  - 4 \Re \sum_i p_i \bigg\langle w_i^{(k)} \bigg| w_i \bigg\rangle \bigg\langle w_i \bigg| w_i^{(l)} \bigg\rangle\\
&=& \sum_i \frac{1}{p_i} \bigg(\frac{\partial p_i}{\partial \theta^j} \bigg)\bigg(\frac{\partial p_i}{\partial \theta^k} \bigg)  + 4 \Re \sum_{i < j} (p_i + p_j ) \bigg\langle w_i^{(k)} \bigg| w_j \bigg\rangle \bigg\langle w_j \bigg| w_i^{(l)} \bigg\rangle.
\end{eqnarray*}

 \begin{lemma}
The $C_L$ quantum information is less than or equal to the SM quantum information $C_\Upsilon$ for multi-parameter families of states, i.e.
\begin{equation}
C_L(\theta) \leq C_{\Upsilon}(\theta).
\label{mulithc}
\end{equation}
\label{lm:4}
\end{lemma}
\proof
Equation (\ref{mulithc}) is equivalent to 
\begin{equation}
 v^T   C_L(\theta)  v \leq  v^T  C_{\Upsilon}(\theta) v ,
\label{eq:cCc}
\end{equation}
for all $v \in \mathbb{R}^p$. To prove  (\ref{mulithc}) we choose suitable one-parameter families of states and use Lemma \ref{lem.clleqcups1}. For given $\theta$ and $ v$ in $\mathbb{R}^p$, consider the set of one-parameter states 
\begin{equation*}
 \rho( \theta + t v) = \sum_{k=1}^d p_k(\theta + t v)| w_k(\theta + t v) \rangle \langle w_k(\theta + t v) |, \qquad t \in \mathbb{R}.
\end{equation*}
Now,
\begin{eqnarray}
\frac{d}{d t}  p_{k} (\theta + t  v  ) &=& \sum_l  \frac{\partial p_{k} (\theta)}{\partial \theta^l} v^l + O(t), 
\label{eq:multip1}\\
\frac{d}{d t}  | w_{k} (\theta + t  v  ) \rangle &=& \sum_l   \bigg|w_k(\theta)^{(l)} \bigg\rangle v^l + O(t),  \quad \bigg|w_k(\theta)^{(l)} \bigg\rangle = \frac{\partial}{\partial \theta^l}   | w_k(\theta) \rangle, 
\label{eq:multip2}
\end{eqnarray}
where $ v^l$ is the $l$th component of the vector $v$. We prove equations (\ref{eq:multip1}) and (\ref{eq:multip2}) in Appendix A.
From Lemma \ref{lem.clleqcups1} we know that $  C_L(t) \leq C_{\Upsilon}(t)$, i.e.
\begin{eqnarray*}
\sum_i \frac{1}{p_i(\theta +tv)} \bigg(\frac{d p_i}{d t} \bigg)^2 &+& 
4 \sum_{j<k} (p_j(\theta + tv) + p_k(\theta + tv))  \bigg| \bigg\langle \frac{d w_j}{dt} \bigg| w_k \bigg\rangle \bigg|^2\\
 \leq \sum_i \frac{1}{p_i(\theta +tv)} \bigg(\frac{d p_i}{d t} \bigg)^2 &+& 
4 \sum_{j<k} (p_j(\theta + tv) + p_k(\theta + tv))  \bigg| \bigg\langle \frac{d w_j}{dt}\bigg| w_k \bigg\rangle \bigg|^2\\
&+& 4 \sum_i p_i(\theta) \bigg| \bigg\langle \frac{d w_i}{dt} \bigg| w_i \bigg\rangle \bigg|^2 .
\label{eq:34}
\end{eqnarray*}
Using (\ref{eq:multip1}) and (\ref{eq:multip2}) and evaluating at $t=0$  gives
\begin{eqnarray*}
\sum_{m,n} v^m v^n \left( \sum_i \frac{1}{p_i} \bigg(\frac{\partial p_i}{\partial \theta^m} \bigg)\bigg(\frac{\partial p_i}{\partial \theta^n} \bigg)  + 4 \sum_{i < j} (p_i + p_j ) \langle w_i^{(m)} | w_j \rangle \langle w_j | w_i^{(n)} \rangle \nonumber \right. \\
\leq \sum_{r,s} v^r v^s \left( \sum_i \frac{1}{p_i} \bigg(\frac{\partial p_i}{\partial \theta^r} \bigg)\bigg(\frac{\partial p_i}{\partial \theta^s} \bigg)  + 4 \sum_{i < j} (p_i + p_j ) \langle w_i^{(r)} | w_j \rangle \langle w_j | w_i^{(s)} \rangle \right)\\
+ \left. 4  \sum_i p_i \langle w_i^{(m)} | w_i \rangle \langle w_i | w_i^{(n)} \rangle \right) .
\label{eq:34b}
\end{eqnarray*}
We can rewrite this as
\begin{equation*}
\sum_{m,n} v^m v^n C_L(\theta)_{mn} \leq \sum_{r,s} v^r v^s C_\Upsilon (\theta)_{rs}. 
\end{equation*}
This is equivalent to (\ref{eq:cCc}). Since this holds for all $v$ in $\mathbb{R}^p$, we have (\ref{mulithc}). 

   \begin{lemma}
Equality holds in (\ref{mulithc}) for families of states (\ref{eq.statesw}) if and only if they satisfy
\begin{eqnarray}
 \langle w_i^{(m)} | w_i \rangle  = 0, \quad \forall m,i, \quad   p_i > 0.
\label{eq:unitcond3}
\end{eqnarray}
\label{eq:equalitym}
\end{lemma}
\proof
Equality in (\ref{mulithc}) is equivalent to 
\begin{equation}
 v^T   C_L(\theta)  v =  v^T  C_{\Upsilon}(\theta) v ,
\label{eq:cCc2}
\end{equation}
for all $v \in \mathbb{R}^p$.  From the proof of Lemma \ref{lm:4} we see that for (\ref{eq:cCc2}) to be satisfied for all $v \in \mathbb{R}^p$, we require that, for one-parameter families of states  $\rho(\theta + tv)$, for given $\theta$ and $v \in \mathbb{R}^p$, we have $C_L(t) |_{t=0} =  C_{\Upsilon}(t) |_{t=0}$. From Lemma \ref{lem.clleqcups1} this is possible if and only if the channel satisfies (\ref{eq.eq1pclcu}) at the point $t=0$. This condition is equal to
\begin{equation*}
\left. \left. \sum_i p_i(t) \left| \left\langle \frac{d w_i}{dt} \right| w_i \right\rangle \right|^2 \right|_{t=0} = 0. 
\end{equation*}
Using (\ref{eq:multip2}) this condition can be rewritten  as 
  \begin{eqnarray}
\sum_{l=1}^{m} v^m v^n   \sum_i p_i \langle w_i^{(m)} | w_i \rangle \langle w_i | w_i^{(n)} \rangle = 0, \quad \forall m,n.
\label{eq:unitcond2}
\end{eqnarray}
Condition (\ref{eq:unitcond2}) holds for all $v$ if and only if (\ref{eq:unitcond3}) is satisfied.

 \begin{lemma}
The SLD quantum information is less than or equal to the $C_L$ quantum information for multi-parameter families of states, i.e.
\begin{equation}
H_{SLD}(\theta) \leq C_L(\theta).
\label{mulithchcl}
\end{equation}
\label{lm:5hc}
\end{lemma}
\proof
This follows from Lemma \ref{lem.clleqh} in  the same way as Lemma \ref{lm:4} follows from Lemma \ref{lem.clleqcups1}.

\begin{lemma}
Equality holds in (\ref{mulithchcl}) for families of states (\ref{eq.statesw}) if and only if they satisfy
\begin{eqnarray*}
 \langle w_j^{(m)} | w_k \rangle = 0, \quad \forall m, \quad j \neq k, p_j,p_k > 0.
\label{eq:unitcond3b}
\end{eqnarray*}
\label{eq:equalitymphc}
\end{lemma}

\proof
This follows from Lemma \ref{eq.eq1phcl} in  the same way as Lemma \ref{eq:equalitym} follows from Lemma \ref{lem.clleqcups1}.

\section{Is $C_L$ the minimum among $C_\Upsilon$? }
In Example 1, we showed that for $C_\Upsilon$, different choices of eigenvectors of $\rho(\theta)$ result in completely different metrics. If $C_\Upsilon$ is to be an upper bound on Fisher information, it seems sensible to choose the minimum among possible values of $C_\Upsilon$.

We now show that in the one-parameter case $C_L$ is the minimum among $C_\Upsilon$. To do this, we need to show that there exists a choice of phase shift, such that (\ref{eq.eq1pclcu}) is satisfied. 

\subsection{One-parameter case}
Given a family of states $\rho_\theta = \sum_i p_i (\theta) | w_i(\theta) \rangle \langle w_i(\theta) |$,
a phase change of the eigenvectors $| w_1(\theta) \rangle, \dots, | w_d(\theta) \rangle$ sends these vectors to $| v_1(\theta) \rangle, \dots, | v_d(\theta) \rangle$, where $| v_j (\theta) \rangle = \exp(i \alpha_j(\theta)) | w_j (\theta) \rangle$ for some real-valued functions $\alpha_1, \dots, \alpha_d$. The density matrix $\rho_\theta$ is unchanged.

Now
\begin{equation*}
\frac{d}{d \theta} | v_k (\theta) \rangle = i \frac{d \alpha_k}{d \theta} \exp(i \alpha_k(\theta)) | w_k (\theta) \rangle +  \exp(i \alpha_k(\theta)) \frac{d}{d \theta} | w_k(\theta) \rangle
\end{equation*}
and hence 
\begin{equation*}
\sum_k p_k | \langle v_k' | v_k \rangle |^2 = \sum_k p_k \bigg| -i \frac{d \alpha_k}{d \theta} +  \langle w_k' | w_k \rangle \bigg|^2.
\end{equation*}
 There are an infinite number of choices of phase shift $\alpha_k$ resulting in an infinite number of different metrics $C_{\Upsilon}(\theta)$. Hence, for a given family of states, the SM quantum information   gives rise to an infinite number of different metrics.
If we choose
\begin{equation*}
\alpha_k(\theta) = - i  \int_{\theta_0}^\theta \langle w_k'(\phi) | w_k(\phi)  \rangle d\phi
\end{equation*}
then (\ref{eq.eq1pclcu}) is satisfied.  Since $ \langle w_k' | w_k \rangle$ is purely imaginary,  $\alpha_k$ is real. Thus for this choice of phase shift, $C_L(\theta) = C_{\Upsilon}(\theta)$. 
From (\ref{eq.clecups}) we see that this choice of phase shift gives us the minimum value of $C_{\Upsilon}(\theta)$ for the family of states $\rho(\theta)$. 

Thus, in the one-parameter case $C_L$ is the minimum among $C_\Upsilon$. We are better to use $C_L$ than $C_\Upsilon$, as the former is a more sensible metric.

\subsection{Multi-parameter case}
We now show that, in general, $C_L$ is not the minimum among $C_\Upsilon$ in the multi-parameter case. A phase change of the eigenvectors $| w_1(\theta) \rangle, \dots, | w_d(\theta) \rangle$ sends these vectors to $| v_1(\theta) \rangle, \dots, | v_d(\theta) \rangle$, where $| v_j (\theta) \rangle = \exp(i \alpha_j(\theta)) | w_j (\theta) \rangle$ for some real-valued functions $\alpha_1, \dots, \alpha_d$. In this case $\theta = (\theta^1, \dots, \theta^p)$.

For equality in (\ref{mulithc}) we require that (\ref{eq:unitcond3}) is satisfied. Now,
\begin{equation*}
\frac{\partial}{\partial \theta^m} | v_j (\theta) \rangle = i \frac{\partial \alpha_j}{\partial \theta^m} \exp(i \alpha_j(\theta)) | w_j (\theta) \rangle +  \exp(i \alpha_j(\theta)) \frac{\partial}{\partial \theta^m} | w_k(\theta) \rangle
\end{equation*}
and hence 
\begin{equation*}
\sum_j p_j  \langle v_j^{(m)} | v_j \rangle \langle v_j | v_j^{(n)}  \rangle  = \sum_j p_j \left( -  i \frac{\partial \alpha_j}{\partial \theta^m} +  \langle w_j^{(m)} | w_j \rangle \right)\left(  i \frac{\partial \alpha_j}{\partial \theta^n} +  \langle w_j | w_j^{(n)} \rangle \right).
\end{equation*}
For this term to be zero, we require that
\begin{equation*}
i \frac{\partial \alpha_j}{\partial \theta^m} = \bigg\langle \frac{\partial w_j}{\partial \theta^m}\bigg| w_j \bigg\rangle \in i \mathbb{R} \qquad \forall j,m.
\end{equation*}
This is solvable if and only if
\begin{equation*}
\frac{\partial^2 \alpha_j}{\partial \theta^k \partial \theta^l} = \frac{\partial^2 \alpha_j}{\partial \theta^l \partial \theta^k} \quad \forall j,k,l. 
\end{equation*}
This is equivalent to
\begin{equation*}
\frac{\partial}{\partial \theta^k} \bigg\langle \frac{\partial w_j}{\partial \theta^l}\bigg| w_j \bigg\rangle = \frac{\partial}{\partial \theta^l} \bigg\langle \frac{\partial w_j}{\partial \theta^k}\bigg| w_j \bigg\rangle \quad \forall j,k,l ,
\end{equation*}
which is equivalent to 
\begin{equation*}
 \bigg\langle \frac{\partial^2 w_j}{\partial \theta^k \partial \theta^l}\bigg| w_j \bigg\rangle +  \bigg\langle \frac{\partial w_j}{\partial  \theta^l}\bigg| \frac{ \partial w_j}{\partial \theta^k} \bigg\rangle= \bigg\langle \frac{\partial^2 w_j}{\partial \theta^l \partial \theta^k} \bigg| w_j \bigg\rangle + \bigg\langle \frac{\partial w_j}{\partial \theta^k}\bigg| \frac{\partial w_j}{\partial \theta^l} \bigg\rangle \quad \forall j,k,l. 
\end{equation*}
Since $| w_j \rangle$ is continuously differentiable, 
\begin{equation*}
 \bigg\langle \frac{\partial^2 w_j}{\partial \theta^k \partial \theta^l}\bigg| w_j \bigg\rangle = \bigg\langle \frac{\partial^2 w_j}{\partial \theta^l \partial \theta^k} \bigg| w_j \bigg\rangle \quad \forall j,k,l ,
 \end{equation*}
 and hence we require that
  \begin{equation*}
 \bigg\langle \frac{\partial w_j}{\partial  \theta^l}\bigg| \frac{ \partial w_j}{\partial \theta^k} \bigg\rangle= \bigg\langle \frac{\partial w_j}{\partial \theta^k}\bigg| \frac{\partial w_j}{\partial \theta^l} \bigg\rangle \qquad \forall j,k,l.
\end{equation*}
This is satisfied if and only if
\begin{equation}
 \bigg\langle \frac{\partial w_j}{\partial  \theta^l}\bigg| \frac{ \partial w_j}{\partial \theta^k} \bigg\rangle \in \mathbb{R} \qquad \forall j,k,l.
 \label{eq:condclcum}
\end{equation}
Hence, for multi-parameter families of states, $C_L$ is not generally the minimum among $C_\Upsilon$.

{\noindent{\bf{Example 2}} 
\newline
For the family of states given in Example 1, 
\begin{eqnarray*}
 \bigg\langle \frac{\partial v_1}{\partial  \theta}\bigg| \frac{ \partial v_1}{\partial \phi} \bigg\rangle = \frac{i}{2} \sin(\theta/2) \cos(\theta/2),\\
  \bigg\langle \frac{\partial v_2}{\partial  \theta}\bigg| \frac{ \partial v_2}{\partial \phi} \bigg\rangle = \frac{-i}{2} \sin(\theta/2) \cos(\theta/2).
\end{eqnarray*}
Since (\ref{eq:condclcum}) is not satisfied,  $C_L$ is not the minimum among $C_\Upsilon$, for this family of states.

 \section{A note on $C_L$ }
Here, we show that $C_L$ can be rewritten in terms of the classical Fisher information and the SLD quantum information. 
For families of states (\ref{eq.statesw}) 
\begin{eqnarray*}
C_L &=& C_{\Upsilon} - 4 \sum_i p_i | \langle w_i' | w_i \rangle |^2 \\
&=& \sum_i \frac{1}{p_i} \bigg(\frac{d p_i}{d \theta} \bigg)^2 + 4 \sum_i ( p_i \langle w_i ' | w_i ' \rangle - p_i | \langle w_i' | w_i \rangle |^2),\\
&=& \sum_i \frac{1}{p_i} \bigg(\frac{d p_i}{d \theta} \bigg)^2 + \sum_i p_i H_{SLD}(\rho_i(\theta)),
  \label{calebmetric2}
\end{eqnarray*}
where $H_{SLD}(\rho_i(\theta))$ is the SLD quantum information for the pure state $\rho_i(\theta) = | w_i(\theta) \rangle \langle w_i(\theta) |$. This interesting result states that the $C_L$ quantum information is equal to the classical Fisher information of the probability distribution $\{ p_1(\theta),\dots,p_d(\theta) \}$ plus a weighted sum of the SLD quantum informations of the pure states $\rho_i(\theta)$ of which the state $\rho(\theta)$ is a convex mixture. 

Similarly, for a multivariate family of states we can show that
\begin{eqnarray*}
C_L(\theta)_{kl} &=& \sum_i \frac{1}{p_i} \bigg(\frac{\partial p_i}{\partial \theta^k} \bigg)\bigg(\frac{\partial p_i}{\partial \theta^l} \bigg) + \sum_i p_i H_{SLD}(\rho_i(\theta))_{kl},
\end{eqnarray*}
where $H_{SLD}(\rho_i(\theta))_{kl}$ is the $(k,l)$th entry of the SLD quantum information for $\rho_i(\theta)$.

\section*{Conclusion}
The SM quantum information $C_\Upsilon$ is not a well-defined metric.  We have defined a new quantum information $C_L$ from $C_\Upsilon$ which is a well-defined metric: invariant but not monotone. We have given relationships between $C_L$, $C_\Upsilon$ and the SLD quantum information $H_{SLD}$. 

\ack
I am very grateful to Peter Jupp for numerous helpful comments.
This work was supported by the EPSRC.

\appendix

\section*{Appendix A}
\setcounter{section}{1}
Here, we prove that 
\begin{equation}
\frac{d }{d t}  p_{k} (\theta + t v ) = \sum_l  p_{k} (\theta)^{(l)} v^l + O(t), \qquad p_{k} (\theta)^{(l)} = \frac{\partial p_{k} (\theta)}{\partial \theta^l}. 
\label{eq:multip1x}
\end{equation}
Let us introduce the vector $\phi(t) = \theta + t v $,  with components $\phi^l = \theta^l + t  v ^l $. Using the chain rule to differentiate $p_{k} (\phi(t) )$, we get
\begin{equation}
\frac{d}{d t} p_{k} (\phi(t)) = \sum_l  \frac{ \partial p_{k} (\phi)}{\partial \phi^l} \frac{ \partial \phi^l}{\partial t} .
\label{eq:appca8}
\end{equation}
Now,
\begin{eqnarray*}
  \frac{ \partial p_{k} (\phi)}{\partial \phi^l}  &=&  \left. \frac{ \partial p_{k} (\phi)}{\partial \phi^l}  \right|_{t=0}+ O(t) = \frac{ \partial p_{k} (\theta)}{\partial \theta^l} + O(t) ,\\
  \frac{ \partial \phi^l}{\partial t}  &=& v^l.
  \end{eqnarray*}
Substituting these back into (\ref{eq:appca8}) gives (\ref{eq:multip1x}). In a similar way we get 

\begin{equation*}
\frac{d }{d t}  | w_{k} (\theta + t v ) \rangle = \sum_l  | w_{k} (\theta)^{(l)} \rangle v^l + O(t). 
\label{eq:w}
\end{equation*}

\end{document}